\documentclass[reprint,superscriptaddress,showpacs,amsmath,amssymb,aps,prl]{revtex4-1}
\usepackage{graphicx}
\usepackage{dcolumn}
\usepackage{bm}
\usepackage[utf8]{inputenc}
\usepackage[frenchb]{babel}
\usepackage{braket}
\usepackage[T1]{fontenc}

\begin{document}

\preprint{APS/123-QED}

\title{Coupled oscillator model of a trapped Fermi gas at the BEC-BCS crossover}

\author{S. V. Andreev}
\email[Electronic adress: ]{Serguey.Andreev@gmail.com}
\affiliation{Physikalisches Institut, Albert-Ludwigs-Universit\"at Freiburg, Hermann-Herder-Strasse 3, 79104 Freiburg, Germany}

\date{\today}

\begin{abstract}

We address theoretically the puzzling discontinuity of the radial quadrupole mode frequency observed in a trapped Fermi gas across the BEC-BCS crossover. We apply the scaling transformation to a two-channel model of a resonant Fermi superfluid and argue that the frequency downshift in the crossover region is due to Feshbach coupling of the molecular Bose-Einstein condensate (BEC) to the surrounding Fermi sea. The Bose and Fermi components of the gas act as coupled macroscopic oscillators. The frequency jump corresponds to the point where the closed-channel molecules are entirely converted into the Fermi sea. This implies linear scaling of the "critical" detuning between the scattering channels with the Fermi energy, which can be readily verified in an experiment.   

\end{abstract}


\maketitle

In ultra-cold atomic gases, collective oscillations are key observables that can be measured with high precision \cite{PitaevskiiBook}. Such measurements offer a unique opportunity to test subtile aspects of the theory. For fermions, the central issue has been the crossover from a weakly-paired Bardeen-Cooper-Schrieffer (BCS) superfluid to a Bose-Einstein condensate (BEC) of tightly bound molecules upon increasing the strength of two-body attraction \cite{Gurarie2007, Giorgini2008, Andreev2022F} \footnote{A broader context may also include high-temperature superconductivity \cite{Chen2005}, nuclear matter \cite{Strinati2018} and stars \cite{Pethick2017}, as well as exciton-polaritons in semiconductors \cite{Edelman2017}}. The attractive interaction of atoms can be tuned at will by means of the Feshbach resonance \cite{Feshbach}. A widely held conviction is that BEC-BCS crossover connects relevant characteristics of an equilibrium system in a continuous manner. In particular, the spectrum of elementary excitations has been expected to evolve smoothly between the predictions of the hydrodynamic theory of superfluids (BEC side) and the collisionless limit (BCS side) \cite{PitaevskiiBook, Combescot2006}. The series of experimental studies outlined below has cast doubt on this belief and posed a challenge to the theory.

Atomic clouds were prepared in axisymmetric harmonic traps in the elongated ($\omega_z\ll\omega_r$) \cite{Altmeyer2007} and flat ($\omega_z\gg\omega_r$) \cite{Vogt2012} geometries. The low-energy collective oscillations in this case take the form of discretized normal modes classified by the projection $l_z$ of the angular momentum on the symmetry axis ($z$-axis). Of particular interest is the quadrupole mode ($l_z=\pm 2$) which corresponds to shape oscillations of the cloud and does not depend on the equation of state \cite{PitaevskiiBook}. In the collisionless limit this mode is analogous to Landau's zero sound in a uniform Fermi liquid \cite{Landau1957}. At zero temperature the frequency of the collisionless quadrupole mode is predicted to be $\omega_q=2\omega_r$ \cite{PitaevskiiBook}, where $\omega_r$ is the radial trapping frequency. In the hydrodynamic regime (either irrotational or classical), one has $\omega_q=\sqrt{2}\omega_r$ \cite{Stringari1996, Pitaevskii1998}. Contrary to the aforementioned expectations, the experiments \cite{Altmeyer2007, Vogt2012} have revealed an abrupt jump between these two values as one tunes the strength of attraction and measures a sequence of equilibrium states of the cloud over the entire crossover. Despite several theoretical attempts \cite{Zhou2008, Urban2011, Urban2013, Bruun2013, Dong2015}, the origin of this discontinuity, as well as the concurrent downshift of $\omega_q$ upon entering the crossover from the BEC side, has not been understood.

In this work, we provide an explanation of the observed behaviour in the frame of a two-channel model of a resonant Fermi superfluid \cite{Gurarie2007, Andreev2022F}. The model establishes reference points for the BEC-BCS crossover. From the BEC side, the crossover begins at the two-body unitarity. As one increases the energy of the closed-channel molecule (resonance) with respect to the open channel, the molecular condensate dissociates into fermions which form a BCS ground state. The crossover terminates at the point where the molecules are entirely converted into the fermions: beyond this point, the boson population is exponentially small. Although there is a unique broken $U(1)$ symmetry associated with conservation of the total number of particles over the entire crossover, the model inherently incorporates a fundamental difference between the BEC and BCS condensates. What makes the difference is \textit{the non-linearity}: whereas Cooper pairs do not interact, the molecules behave as weakly-repulsive bosons. We argue that the respective solutions for $\omega_q$ in BEC and BCS phases stem from different dynamical scaling and cannot be connected in a continuous fashion. In the crossover region, the BEC and BCS condensates constitute two distinct macroscopic oscillators coupled to each other via a coherent Feshbach link. The strength of the coupling increases with spatial overlap between the components and buildup of a Fermi surface, as one moves toward the BCS side.

The model Hamiltonian reads
\begin{equation}
\label{Hamiltonian}
\begin{split}
\hat H(t)=&\int\sum_{\sigma=\uparrow,\downarrow,B}\hat\Psi^\dagger_\sigma(\bm r,t)\left[-\frac{\hbar^2}{2m_\sigma}\Delta+V_\sigma(\bm r,t)\right]\hat\Psi_\sigma(\bm r,t) d\bm r\\
&+\frac{1}{2}\sum_{\sigma,\sigma'}\bar g_{\sigma\sigma'}\int\hat\Psi^{\dag}_{\sigma}\hat\Psi^{\dag}_{\sigma'}\hat\Psi_{\sigma'}\hat\Psi_{\sigma} d\bm r+\bar\delta\int\hat\Psi_{B}^\dagger\hat\Psi_{B}d\bm r\\
&-\alpha\int\hat\Psi^\dagger_\uparrow(\bm r, t)\hat\Psi^\dagger_{\downarrow}(\bm r, t)\hat\Psi_{B}(\bm r, t)d\bm r-\mathrm{H.c.},
\end{split}
\end{equation}
where fermions of equal masses $m_\uparrow=m_\downarrow\equiv m$ are described by the second-quantized fields $\hat\Psi_{\uparrow}(\bm r, t)$ and $\hat\Psi_{\downarrow}(\bm r, t)$, and the field operator $\hat\Psi_{B}(\bm r, t)$ stands for their bosonic molecules with mass $m_B=2m$. We shall neglect thermal and quantum depletion of the molecular condensate and replace $\hat\Psi_{B}(\bm r, t)$ by a classical field $\Psi_{B}(\bm r, t)$. The effective interactions $\bar g_{\sigma\sigma'}=2\pi\hbar^2/m_{\sigma\sigma^\prime}\bar a_{\sigma\sigma^\prime}$ are defined by the corresponding reduced masses $m_{\sigma\sigma^\prime}=m_{\sigma} m_{\sigma^\prime}/(m_{\sigma}+m_{\sigma^\prime})$ and $s$-wave scattering lengths $\bar a_{\sigma\sigma^\prime}$, and include the background attraction between the fermions of opposite spins $\bar g_{\uparrow\downarrow}<0$. We assume, however, that the pairing is dominated by the Feshbach resonance described by the last two terms. Namely, for a singlet pair of fermions in vacuum, the Hamiltonian \eqref{Hamiltonian} yields the scattering length \cite{Andreev2022F} 
\begin{equation}
\label{ScattLengthDelta}
a_{\uparrow\downarrow}=\bar a_{\uparrow\downarrow}-\pi^{-1}(m/\hbar^2)\alpha^2/\delta.
\end{equation}
The parameter $\alpha>0$ is proportional to the microscopic volume of the closed-channel molecule $\upsilon$ and to the Josephson energy associated with the coherent Feshbach coupling (\textit{i. e.}, the hyperfine interaction). We assume $\upsilon^{1/3}\ll \sqrt{\hbar/m\omega_\mathrm{ho}}$, where $\omega_\mathrm{ho}=\sqrt{\omega_r\omega_z}$. The renormalized detuning $\delta=\bar\delta+\delta_\alpha$ is reduced with respect to its bare value $\bar\delta$ by the amount $\delta_\alpha\propto-\alpha^2$. 

In practice, the bare detuning $\bar\delta$ is controlled by the Zeeman splitting between the open and closed channels of the Feshbach resonance. By writing $\delta=\mu_B \mathrm{g} (B-B_0)$ and defining 
\begin{equation}
\label{ResWidth}
\Delta B\equiv m\alpha^2/(\pi\mu_B\mathrm{g}\hbar^2\bar a_{\uparrow\downarrow}),
\end{equation}
we may recast the above expression for $a_{\uparrow\downarrow}$ in the familiar form \cite{PitaevskiiBook}
\begin{equation}
\label{ScattLength}
a_{\uparrow\downarrow}=\bar a_{\uparrow\downarrow}\left(1-\frac{\Delta B}{B-B_0}\right).
\end{equation}
The magnetic field $B=B_0$ corresponds to the unitarity, where the scattering length $a_{\uparrow\downarrow}$ diverges. The formulas \eqref{ResWidth} and \eqref{ScattLength} establish a link between the model and the experiments.

The bare detuning $\bar\delta$ together with the total number of particles
\begin{equation}
\label{Norm}
N=\int(\braket{\hat\Psi_{\uparrow}^\dagger\hat\Psi_{\uparrow}}+\braket{\hat\Psi_{\downarrow}^\dagger\hat\Psi_{\downarrow}}+2|\Psi_{B}|^2)d\bm r
\end{equation}
are the control parameters which define the equilibrium configuration and dynamical properties of the system. For instance, behaviour of the gas in a time-dependent harmonic trap
\begin{equation}
V_\sigma(\bm r, t)=\frac{m_\sigma}{2}[\omega_x(t)^2 x^2+\omega_y(t)^2 y^2+\omega_z(t)^2 z^2]
\end{equation}
is well understood in the limiting cases $\bar\delta<-\delta_\alpha$ (BEC regime) \cite{Pitaevskii1998, Edwards1996, Jin1996, Jochim2003, Kagan1996, Kagan1997} and $\bar\delta\gg 2\mu$ (BCS regime) \cite{Bruun2000, Urban2005}, where $\mu(N)$ is the Fermi energy calculated in that latter limit. The intermediate range $-\delta_\alpha<\bar \delta\lesssim 2\mu$ corresponds to the BEC-BCS crossover regime addressed in this work. 

We shall be interested in the radial quadrupole oscillation of the cloud, which can be triggered, \textit{e. g.}, by a sudden quench of a slightly anizotropic trap $\omega_x(0)\neq\omega_y(0)$ to the axially symmetric configuration $\omega_x(t)=\omega_y(t)\equiv\omega_r$, the latter then being retained at all times $t>0$. To describe the resulting shape oscillations, we perform the scaling transformation \cite{Castin1996, Kagan1996, Kagan1997, Dalfovo1997, Bruun2000}
\begin{equation}
\begin{split}
\hat\Psi_{\uparrow,\downarrow}(\bm r,t)&=\frac{1}{\sqrt{\mathcal{V}_b(t)}}\hat\chi_{\uparrow,\downarrow}[\bm\rho(t),\mathsf{t}(t)]e^{i\Phi_{\uparrow,\downarrow}(\bm r,t)}\\
\Psi_B(\bm r,t)&=\frac{1}{\sqrt{\mathcal{V}_c(t)}}\chi_B[\bm{\varrho}(t),\mathfrak{t}(t)]e^{i\Phi_B(\bm r,t)} 
\end{split}
\end{equation}
with $\rho_i(t)=x_i/b_i(t)$, $\varrho_i(t)=x_i/c_i(t)$, $\mathcal{V}_b(t)=b_x(t)b_y(t)b_z(t)$, $\mathcal{V}_c(t)=c_x(t)c_y(t)c_z(t)$ $\mathsf{t}(t)=\int\limits^{t}\mathcal{V}_b^{-1}(t^\prime)dt^\prime$, $\mathfrak{t}(t)=\int\limits^{t}\mathcal{V}_c^{-1}(t^\prime)dt^\prime$ and the quadratic \textit{ansatzes} for the phases
\begin{equation}
\label{Phases}
\begin{split}
\Phi_\uparrow(\bm{r},t)&=\Phi_\downarrow(\bm{r},t)=\frac{m}{2\hbar}\sum_{i=x,y,z}\frac{\dot b_i}{b_i}x_i^2\\
\Phi_B(\bm{r},t)&=\frac{m}{\hbar}\sum_{i=x,y,z}\frac{\dot c_i}{c_i}x_i^2 \; (\mathrm{mod}\:\pi)
\end{split}
\end{equation}
As a first step, let us neglect the Feshbach coupling between the channels by sending $\alpha\rightarrow 0$. The crossover boundaries in this limit become
\begin{equation}
\label{CrossoverRange}
0<\delta< 2\mu,
\end{equation}
and one has $\delta=\bar\delta$. The equations of motion for the fermionic fields and the molecular condensate order parameter in the new variables read
\begin{widetext}
\begin{subequations}
\label{MotionEqs}
\begin{align}
\label{MotionEqsF}
i\hbar\frac{\partial\hat\chi_{\uparrow,\downarrow}}{\partial \mathsf{t}}&=\left[-\frac{\hbar^2}{2m}\sum_i\frac{\mathcal{V}_b}{b_i^2}\frac{\partial^2}{\partial\rho_i^2}+\frac{m}{2}\sum_i\omega_{0i}^2\rho_i^2+\bar g_{F B}|\chi_B|^2\frac{\mathcal{V}_b}{\mathcal{V}_c}\right]\hat\chi_{\uparrow,\downarrow} \\
\label{MotionEqsB}
i\hbar\frac{\partial \chi_B}{\partial \mathfrak{t}}&=\left[-\frac{\hbar^2}{4m}\sum_i\frac{\mathcal{V}_c}{c_i^2}\frac{\partial^2}{\partial\varrho_i^2}+m\sum_i\omega_{0i}^2\varrho_i^2+\bar g_{F B}\left(\braket{\hat\chi_\uparrow^\dagger\hat\chi_\uparrow}+\braket{\hat\chi_\downarrow^\dagger\hat\chi_\downarrow}\right)\frac{\mathcal{V}_c}{\mathcal{V}_b}+\bar g_{BB}|\chi_B|^2+\delta\right]\chi_B,
\end{align}
\end{subequations}
\end{widetext} 
where one has
\begin{subequations}
\begin{align}
\label{ScalingFermions}
\ddot b_i+\omega_i^2 b_i&=\frac{\omega_{0i}}{\mathcal V_b b_i^2}\\
\label{ScalingBosons} 
\ddot c_i+\omega_i^2 c_i&=\frac{\omega_{0i}}{\mathcal V_c c_i^2}
\end{align}
\end{subequations}
with $\omega_{0i}\equiv\omega_i(0)$. Here and in what follows we write the constant quantities having the dimension of energy by using the corresponding units of time. Thus, the detuning $\delta$ in Eq. \eqref{MotionEqsB} has been rescaled by the factor $\mathcal{V}_c$. We have also introduced the notation $\bar g_{FB}\equiv \bar g_{\uparrow B}=\bar g_{\downarrow B}$ for the effective interaction of fermions with bosons. Finally, in Eq. \eqref{MotionEqsF} we have omitted the terms $\bar g_{\uparrow\downarrow}\braket{\hat\chi_\downarrow \hat\chi_\uparrow}\hat\chi_{\downarrow,\uparrow}^\dagger$ which would be exponentially suppressed in the dilute limit [see Eq. \eqref{AnomalousAverage} below].  

Consider limiting forms of Eqs. \eqref{MotionEqs}. First, on the BCS side one has $|\chi_B|\equiv 0$ and Eq. \eqref{MotionEqsF} becomes the equation of motion of an ideal Fermi gas in a harmonic trap. By writing $\hat\chi_{\uparrow,\downarrow}=\hat\xi_{\uparrow,\downarrow}(\rho_x)\hat\xi_{\uparrow,\downarrow}(\rho_y)\hat\xi_{\uparrow,\downarrow}(\rho_z)$ this equation can be further reduced to three independent equations
\begin{equation}
\label{IdealFermi}
i\hbar\frac{\partial\hat\xi_{\uparrow,\downarrow}}{\partial\tau_i}=\left[-\frac{\hbar^2}{2m}\frac{\partial^2}{\partial\rho_i^2}+V_{\uparrow,\downarrow}^{(i)}(\rho_i)\right]\hat\xi_{\uparrow,\downarrow}(\rho_i)
\end{equation}
with $\tau_i(t)=\int\limits^{t}b_i^{-2}(t^\prime)dt^\prime$, $V_{\uparrow,\downarrow}^{(i)}(\rho_i)=m\omega_{0i}^2\rho_i^2/2$ and
\begin{equation}
\label{IdealFermiScaling}        
\ddot b_i+\omega_i^2 b_i=\frac{\omega_{0i}^2}{b_i^3},
\end{equation}
the latter now replacing the coupled equations \eqref{ScalingFermions}.
Following our protocol for excitation of the quadrupole mode, we write $\omega_{0x}=(1+\mathsf{c})\omega_r$ and $\omega_{0y}=(1-\mathsf{c})\omega_r$, and look for the solutions of Eq. \eqref{IdealFermiScaling} in the form $b_i(t)=1+\delta b_i(t)$ with the initial conditions $b_i(0)=1$ and $\dot b_i(0)=0$. Assuming $\mathsf{c}\ll 1$, we obtain 
\begin{equation}
\begin{split}
\delta b_{x}(t)&=\tfrac{1}{2}\mathsf{c}(1-\cos[\omega_q ^{(\mathrm{BCS})} t])\\
\delta b_{y}(t)&=-\tfrac{1}{2}\mathsf{c}(1-\cos[\omega_q ^{(\mathrm{BCS})} t])
\end{split}
\end{equation}
with
\begin{equation}
\label{BCSresult}
\omega_q^{(\mathrm{BCS})}=2\omega_r.
\end{equation}
Finite molecular density $|\chi_B|^2$ goes as a perturbation to the external harmonic potential and does not prevent Eq. \eqref{MotionEqsF} from factorization. The structure of the final Eq. \eqref{IdealFermi} is preserved. The scaling equation Eq. \eqref{IdealFermiScaling} in the presence of the condensate would take the from of a \textit{damped driven harmonic oscillator}. The solution is an oscillation with the frequency of the driving force. Detailed argument will be provided below.

In the opposite limit of $\braket{\hat\chi_\uparrow^\dagger\hat\chi_\uparrow}=\braket{\hat\chi_\downarrow^\dagger\hat\chi_\downarrow}=0$ (BEC side), one may take advantage of the Thomas-Fermi approximation for the molecular condensate order parameter to obtain a stationary solution of Eq. \eqref{MotionEqsB} in the form $\chi_B=|\chi_B|e^{-i 2\mu\mathfrak{t}}$ and
\begin{equation}
\label{ThomasFermiResult}
g_{BB}|\chi_B(\bm\varrho)|^2=2\mu-\delta-m\sum_i \varrho_i^2\omega_{0i}^2.
\end{equation}
The positive difference $2\mu-\delta>0$ is defined by the normalization condition \eqref{Norm}. Solution of the system of coupled scaling equations \eqref{ScalingBosons}, where we put $c_z(t)\equiv 1$, yields 
\begin{equation}
\label{ScalingSolutionsB}
\begin{split}
\delta c_{x}(t)&=\mathsf{c}(1-\cos[\omega_q ^{(\mathrm{BEC})} t])\\
\delta c_{y}(t)&=-\mathsf{c}(1-\cos[\omega_q ^{(\mathrm{BEC})} t])
\end{split}
\end{equation}
with
\begin{equation}
\label{BECresult}
\omega_q^{(\mathrm{BEC})}=\sqrt{2}\omega_r
\end{equation}
and $\mathsf{c}\ll 1$. 

Formally, the difference between the BEC [Eq. \eqref{BECresult}] and BCS [Eq. \eqref{BCSresult}] results can be traced back to absence of the non-linear term in Eq. \eqref{MotionEqsF}, which allows subsequent factorization of the scaling equations \eqref{ScalingFermions}. We now show that both results hold simultaneously within the crossover range \eqref{CrossoverRange}, where the BEC and BCS condensates coexist. No new oscillation arise and the values of the frequencies $\omega_q^{(\mathrm{BCS})}$ and $\omega_q^{(\mathrm{BEC})}$ remain intact.      

At $0<\delta\ll 2\mu$ we may still use Eq. \eqref{ThomasFermiResult}, which we substitute into Eq. \eqref{MotionEqsF} and obtain the factorized form analogous to Eq. \eqref{IdealFermi}, where now the external harmonic potential $V_i(\rho_i)$ should be substituted by an effective potential 
\begin{equation}
\label{EffectivePotential}
U_{\uparrow,\downarrow}^{(i)}(\rho_i)=\tfrac{1}{2}m\tilde\omega_{0i}^2\rho_i^2+\tfrac{g_{F B}}{g_{BB}}(2\mu-\delta)
\end{equation}
in the region of space where Eq. \eqref{ThomasFermiResult} yields non-zero condensate density. Hence, the problem in this region has been reduced to an ideal Fermi gas residing in a superposition of the external potential and an effective mean-field potential produced by the molecular condensate \cite{Molmer1998}. In 3D one has $g_{FB}/g_{BB}\approx 3$ \cite{Petrov2003, Petrov2004}, so that square of the rescaled frequency $\tilde\omega_{0i}^2=\omega_{0i}^2\left(1-\tfrac{2g_{FB}}{g_{BB}}\right)$
is negative and the effective potential \eqref{EffectivePotential} has the form of an inverted parabola. 

The fermions thus form a shell around the molecular core. The exact form of the fermion density profile can be worked out by using the semiclassical approach \cite{PitaevskiiBook}. Oscillation of the outer part of the shell, which feels only the external potential, is still governed by Eq. \eqref{IdealFermiScaling}. The inner part, which feels the effective potential \eqref{EffectivePotential}, oscillates according to the modified scaling equations
\begin{equation}
\label{IdealFermiScalingEqs}
\begin{split}       
\ddot b_x+\tilde\omega_x^2  b_x&=\frac{\tilde\omega_{0x}^2}{b_x^3}+\mathsf{c}\tfrac{2g_{F B}}{g_{BB}} [\omega_q^{(\mathrm{BEC})}]^2\cos[\omega_q^{(\mathrm{BEC})}t]\\
\ddot b_y+\tilde\omega_y^2  b_y&=\frac{\tilde\omega_{0y}^2}{b_y^3}-\mathsf{c} \tfrac{2g_{F B}}{g_{BB}}[\omega_q^{(\mathrm{BEC})}]^2\cos[\omega_q^{(\mathrm{BEC})}t]
\end{split}
\end{equation}
where $\tilde\omega_i^2=\omega_i^2\left(1-\tfrac{2g_{FB}}{g_{BB}}\right)<0$ and we have used Eq. \eqref{ScalingSolutionsB} for the quadrupole oscillation of the condensate. These are equations of \textit{damped driven harmonic oscillators}. Their solutions are linear superpositions of the transients
\begin{equation}
\begin{split}
\delta b_{x}(t)&=\tfrac{1}{2}\mathsf{c}(1-\exp[-\tilde\omega_q ^{(\mathrm{BCS})} t])\\
\delta b_{y}(t)&=-\tfrac{1}{2}\mathsf{c}(1-\exp[-\tilde\omega_q ^{(\mathrm{BCS})} t]),
\end{split}
\end{equation}
with $\tilde\omega_q ^{(\mathrm{BCS})}=2|\tilde\omega_r|$ and oscillations at the frequency of the driven force $\omega_q ^{(\mathrm{BEC})}$, given by Eq. \eqref{BECresult}. The driven oscillations of the fermion shell are in-phase with the molecular BEC.

\begin{figure}[t]
\centering
\includegraphics[width=1\columnwidth]{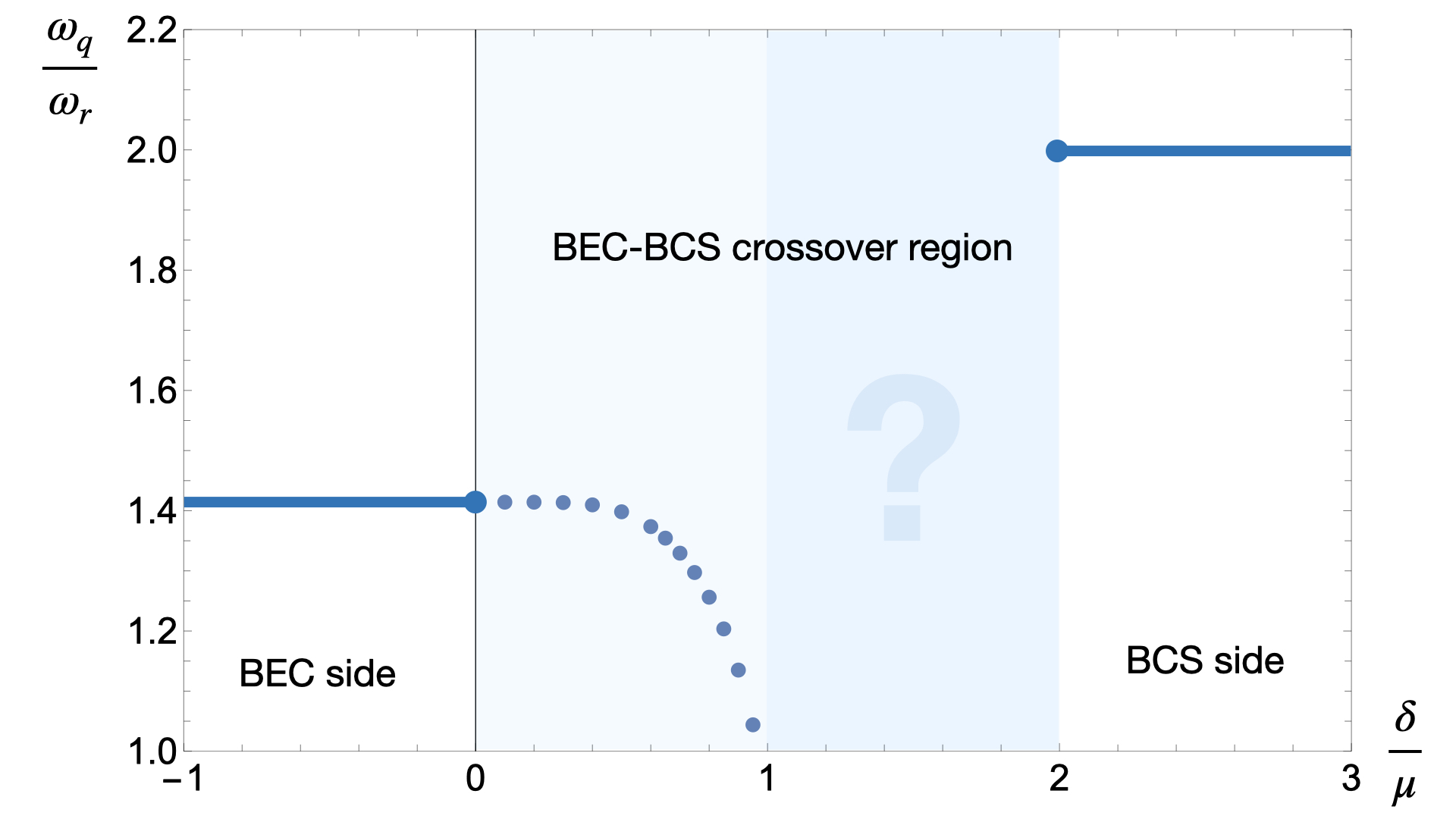}
\caption{Quadrupole oscillation frequency as a function of the detuning. Solid lines are the exact values obtained within the scaling approach in the BEC [Eq. \eqref{BECresult}] and BCS [Eq. \eqref{BCSresult}] limits. The end-points (large filled circles) mark the BEC-BCS crossover (shaded area) boundaries. The vertical line at $\delta=0$ marks the unitarity. The dots represent the perturbative result \eqref{PerturbativeResult}. We have used the parameters of the experiment \cite{Altmeyer2007}. Understanding of behaviour in the doubly-shaded area would require further refinement of the theory.}
\label{Fig1}
\end{figure} 

As the fermion density grows upon increasing $\delta$, the mutual repulsion with the molecular BEC starts to contribute also into Eq. \eqref{MotionEqsB}. However, the corresponding effective potential now has positive curvature and, being approximated by a parabola, yields positive square of the rescaled frequency. The oscillator associated with the molecular BEC, although being driven by the fermionic component in the overlap region, is \textit{undamped} and thus preserves its normal mode \footnote{The rescaled frequency $\tilde\omega_q^{(\mathrm{BEC})}$ would experience a \textit{constant} upshift with respect to the bare value $\omega_q^{(\mathrm{BEC})}$, in qualitative analogy to the recent experimental result \cite{Grimm2019}. However, this effect clearly goes beyond the perturbative treatment at $0<\delta\ll 2\mu$ carried out in this work and by no means affects the ensuing conclusions on the physical origin of the discontinuity.}. We conclude that, within the Thomas-Fermi approximation, there are two independent modes [Eq. \eqref{BCSresult} and Eq. \eqref{BECresult}] which remain intact over the entire crossover.

We now restore the Feshbach coupling $\alpha$ between the channels and treat it as a perturbation to Eqs. \eqref{MotionEqs}. The corresponding corrections to the scaling equations can be worked out by considering dynamics of the average squared radii 
\begin{equation}
\braket{\hat x_{i}^2(t)}_\sigma=b_i^2(t)\braket{\hat x_{i}^2(0)}_\sigma=\frac{1}{N_\sigma(t)}\int  n_\sigma(t) x_i^2(t)d\bm r,
\end{equation}
($\sigma=\uparrow,\downarrow$) and analogously for $c_i^2(t)$ ($\sigma=B$). The full time derivative of $\braket{\hat x_i^2(t)}_\sigma$ would contain corrections to $\dot N_\sigma$ and $\dot n_\sigma$ due to Josephson currents induced by the Feshbach link:
\begin{equation}
\begin{split}
\dot n_{\uparrow,\downarrow}&=\dot n_{\uparrow,\downarrow}^{(0)}+\frac{2\alpha}{\hbar}\sin(\Delta\Phi)|\braket{\hat\Psi_\downarrow\hat\Psi_\uparrow}\Psi_B^\ast|\\
\dot n_{B}&=\dot n_B^{(0)}-\frac{2\alpha}{\hbar}\sin(\Delta\Phi)|\braket{\hat\Psi_\downarrow\hat\Psi_\uparrow}\Psi_B^\ast|,
\end{split}
\end{equation}
where $n_\sigma^{(0)}(t)$ stand for the uncoupled densities and we have defined the relative phase
\begin{equation}
\label{RelativePhase}
\Delta\Phi\equiv\Phi_\uparrow+\Phi_\downarrow-\Phi_B.
\end{equation}
Note, that $\dot N_\sigma^{(0)}=\int \dot n_\sigma^{(0)}d\bm r=0$. By using Eqs. \eqref{Phases}, one may obtain
\begin{subequations}
\label{CoupledEquations}                
\begin{align}
\delta \ddot b_i&=\delta \ddot b_i^{(0)}+\Gamma_{ii}(\delta \ddot b_i-\delta \ddot c_i)+\Gamma_{ij}(\delta \ddot b_j-\delta \ddot c_j)\\
\delta \ddot c_i&=\delta \ddot c_i^{(0)}+\Lambda_{ii}(\delta \ddot c_i-\delta \ddot b_i)+\Lambda_{ij}(\delta \ddot c_j-\delta \ddot b_j),
\end{align}
\end{subequations}
where $j\neq i$ and expressions for the coupling matrix elements may be found in \cite{SI}. By virtue of the cylindrical symmetry of the problem, one has $\Gamma_{xx}=\Gamma_{yy}$, $\Gamma_{xy}=\Gamma_{yx}$ and $\Lambda_{xx}=\Lambda_{yy}$, $\Lambda_{xy}=\Lambda_{yx}$. The zeroth-order quantities $\delta \ddot b_i^{(0)}$ and $\delta \ddot c_i^{(0)}$ obey the uncoupled scaling equations derived above for $\alpha=0$.

Thus constructed system of coupled differential equations on the scaling parameters does no longer admit a global quadrupole solution which would be compatible with the previously used initial conditions, \textit{i. e.}, $\delta b_i(0)=0$, $\delta \dot b_i(0)=0$ \textit{and, simultaneously,} $\delta c_i(0)=0$, $\delta \dot c_i(0)=0$. We, therefore, relax our statement of the problem by assuming that only the majority component (the molecular BEC) dynamics is subjected to the initial conditions $\delta c_i(0)=0$, $\delta \dot c_i(0)=0$, whereas the phase and amplitude of the fermion oscillation is defined entirely by the coupling. This is consistent with our above conclusion on the driven nature of the fermion oscillator. Formally, this amounts to reducing the system of four coupled equations \eqref{CoupledEquations} to just two coupled equations for the differences $\delta \ddot b\equiv \delta \ddot b_x-\delta \ddot b_y$ and $\delta \ddot  c\equiv  \delta \ddot c_x-\delta \ddot c_y$. By taking $\delta \ddot b=\mathsf{b}\omega_q^2\cos(\omega_q t)$, $\delta \ddot c=\mathsf{c}\omega_q^2\cos(\omega_q t)$ and assuming $\omega_q=\sqrt{2}\omega_r+\delta\omega_q$ with $\delta\omega_q\ll \sqrt{2}\omega_r$, we arrive at the eigenvalue problem
\begin{equation}
\label{EigenvalueProblem}
\left |
\begin{array}{cc}
 2\omega_r^2[1+\Gamma(\delta)]-\omega_q^2 &2\omega_r^2\Gamma(\delta)\\
 2\omega_r^2\Lambda(\delta) &2\omega_r^2[1+\Lambda(\delta)]-\omega_q^2 
 \end{array}\right |=0,
 \end{equation}
 where 
 \begin{equation}
 \label{Coupling}
 \Gamma(\delta)=\frac{\alpha m}{2\hbar^2}\cos(\Delta\Phi)\frac{\int\limits_{-\infty}^{+\infty}d\rho_z\int\limits_0^{+\infty}|\braket{\hat\chi_\downarrow\hat\chi_\uparrow}\chi_B^\ast|\rho^5 d\rho}{\int\limits_{-\infty}^{+\infty}d\rho_z\int\limits_0^{+\infty}\braket{\hat\chi_\uparrow^\dagger\hat\chi_\uparrow}\rho^3 d\rho}
 \end{equation}
 and an analogous expression for $\Lambda(\delta)$ is obtained by replacing the rescaled coordinates ($\bm\rho$ by $\bm\varrho$) and the density profile [$\braket{\hat\chi_\uparrow^\dagger(\bm\rho)\hat\chi_\uparrow(\bm\rho)}$ by $|\chi_B(\bm\varrho)|^2$] in the denominator. 
 
 The secular equation \eqref{EigenvalueProblem} has two solutions: $\omega_q=\sqrt{2}\omega_r$ and
 \begin{equation}
 \label{PerturbativeResult}
 \omega_q^{(\alpha)}(\delta)=\omega_r\sqrt{2[1+\Gamma(\delta)+\Lambda(\delta)]}.
 \end{equation}
The former eigenvalue corresponds to an out-of-phase oscillation of the components, and the latter one corresponds to an in-phase oscillation with slightly different amplitudes. Which one of the two solutions has the lowest energy depends on the static value of the relative phase $\Delta\Phi$ [Eq. \eqref{RelativePhase}] between the BEC and BCS condensates. In a ground state one would expect $\Delta\Phi=0$, which allows maximum energy gain $E_\alpha(\delta)$ due to the Feshbach coupling [the last term in Eq. \eqref{Hamiltonian}]. In the excited state under consideration, however, one can overweight that gain by lowering the macroscopic oscillation energy $E_q(\delta)=N\hbar\omega_q(\delta)$. Namely, provided $|E_q(\delta)-E_q(0)|>|E_\alpha(\delta)|$, the condensates will tend to lock their relative phase at $\Delta\Phi=\pi$. The second branch given by Eq. \eqref{PerturbativeResult} then would have the lowest energy and would exhibit monotonous downshift with increasing detuning $\delta$.    
        
We evaluate Eq. \eqref{Coupling} by using the local density approximation (LDA) for the relevant averages \cite{PitaevskiiBook}. Thus, the BEC condensate density is given by Eq. \eqref{ThomasFermiResult} and the anomalous average due to the background attraction between the fermions with opposite spins may be estimated as
\begin{equation}
\label{AnomalousAverage}
 |\braket{\hat\chi_\downarrow(\bm\rho)\hat\chi_\uparrow(\bm\rho)}|=\frac{\epsilon_F(\bm \rho)}{|\bar g_{\uparrow\downarrow}|}\exp\left({-\frac{\pi}{2 k_F(\bm\rho)|\bar a_{\uparrow\downarrow}|}}\right),
\end{equation}
where $\epsilon_F(\bm \rho)\equiv\hbar^2 k_F^2(\bm\rho)/2m$ is the local value of the Fermi energy in the fermion shell, which is related to the local fermion density by
\begin{equation}
(2\pi)^3 \braket{\hat\chi_\uparrow^\dagger(\bm\rho)\hat\chi_\uparrow(\bm\rho)}=\frac{4}{3}\pi\left[\frac{2m\epsilon_F(\bm \rho)}{\hbar^2}\right]^{3/2}.
\end{equation}
The local Fermi energy $\epsilon_F(\bm \rho)$ does not include the external trapping potential and reaches its maximum value $\delta/2$ at the boundary of the molecular BEC. Hence, increasing the detuning $\delta$ yields exponential growth of the anomalous average. Together with the increase of the spatial overlap between the BEC and BCS condensates, this effect contributes to growth of the absolute values of the coupling parameters $\Gamma(\delta)$ and $\Lambda(\delta)$. 

The results of calculation for the parameters of Ref. \cite{Altmeyer2007} are presented in Fig. \ref{Fig1}. Good quantitative agreement with the experiment justifies \textit{a posteriori} our perturbative approach. We check that the solution \eqref{PerturbativeResult} does indeed correspond to the lowest energy at small detuning $0<\delta\ll 2\mu$  \cite{SI}. Understanding of the behaviour in the intermediate range $0\ll \delta < 2\mu$ would require further refinement of the theory [as to include, \textit{e.g}, the quantum-pressure corrections to the Thomas-Fermi expressions].  

To conclude, we have demonstrated that the discontinuity in the radial quadrupole oscillation of a Fermi gas across the BEC-BCS crossover reflects different interaction properties of the condensates. The tightly bound molecules experience weak two-body repulsion, whereas the Cooper pairs do not interact with each other. In the crossover region the fermion shell featuring a residual BCS condensate and the molecular BEC core represent two coupled macroscopic oscillators. The coupling splits the energies of the two modes, with the lower frequency undergoing an increasing downshift upon moving toward the BCS side. The molecular BEC being over, the frequency experiences an abrupt jump toward its BCS value, prescribed by the dynamics of an ideal (Fermi) gas.  Importantly, the model predicts simple scaling of the "critical" detuning $\delta_c$ with the total number of particles $N$: $\delta_c(N)=2\mu(N)$. By using Eq. \eqref{ScattLengthDelta} this may be expressed in terms of the scattering length $a_{\uparrow\downarrow}$. Besides shift of $\delta_c(N)$ toward $0$, we would also expect an upshift of the Thomas-Fermi result \eqref{BECresult} toward the ultimate (an ideal gas) value $2\omega_r$ upon reducing $N$, consistently with the previous studies \cite{Edwards1996, Jin1996, Stringari1996}.

I thank Philipp Lunt, Johannes Reiter and Selim Jochim for introducing me into trapped fermions, drawing my attention to the experiment \cite{Altmeyer2007}, sharing their own recent experimental results on mesoscopic traps and numerous stimulating discussions. The work has been supported by the BW-Stiftung through Grant No. QT-9 NEF2D.

\bibliography{Bibliography}
\end{document}


\title{Supplemental Information for the manuscript \\"Coupled oscillator model of a trapped Fermi gas at the BEC-BCS crossover"}

\author{S. V. Andreev}
\email[Electronic adress: ]{Serguey.Andreev@gmail.com}
\affiliation{Physikalisches Institut, Albert-Ludwigs-Universit\"at Freiburg, Hermann-Herder-Strasse 3, 79104 Freiburg, Germany}

\maketitle
\section{Density profile of the fermion shell in the crossover region}
\label{section1}

\begin{figure}[t]
\centering
\includegraphics[width=0.8\columnwidth]{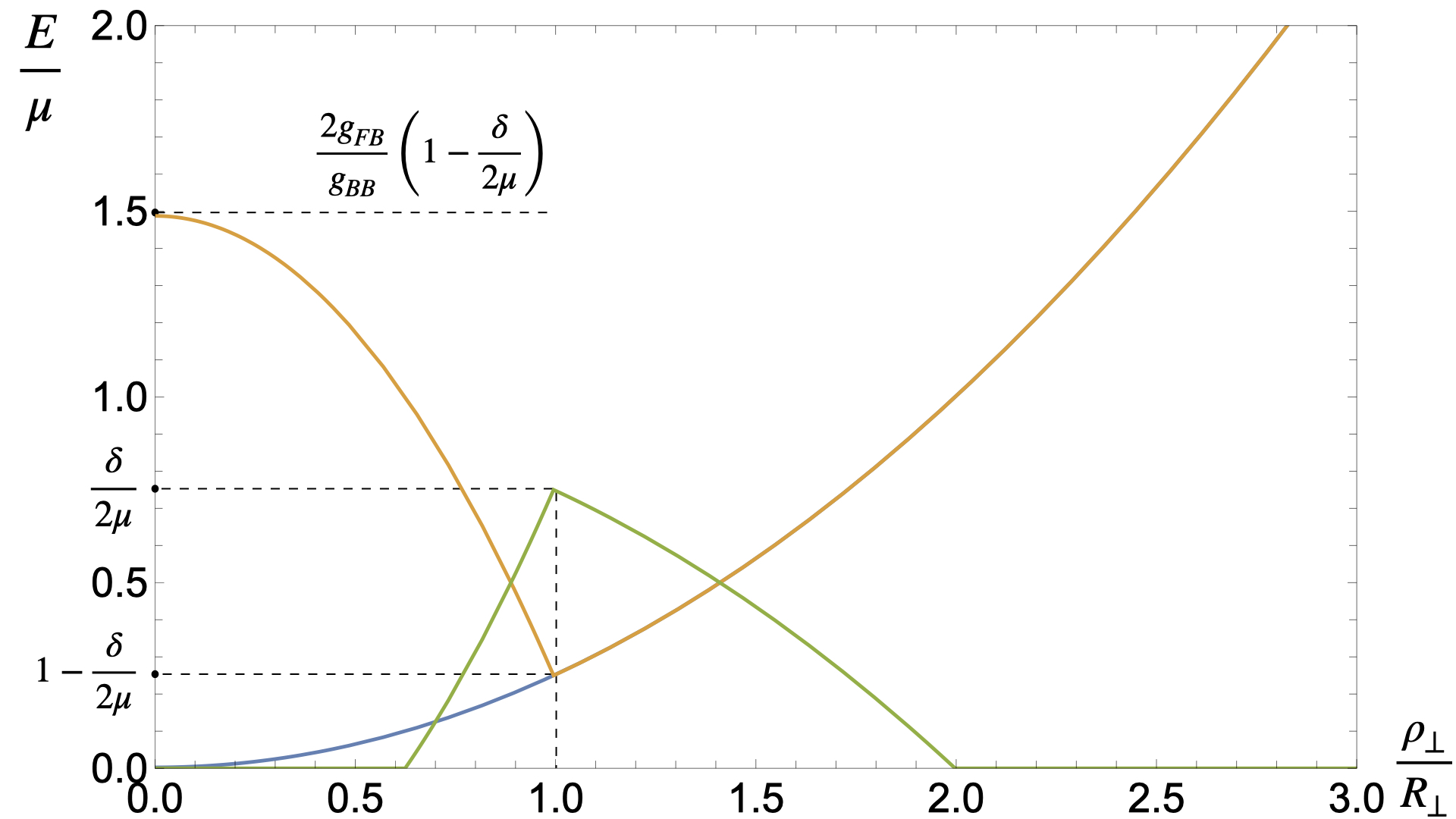}
\caption{Spatial distribution of the local Fermi energy $\epsilon_F(\bm{\tilde\rho})$ (green line) in the radial direction. We take $\tilde\rho_z=0$, $\delta=1.5\mu$ and $g_{FB}/g_{BB}=3$ \cite{Petrov2003, Petrov2004}, and use Eq. \eqref{FermiEnergyInside} and Eq. \eqref{FermiEnergyOutside}. Brown line depicts the effective potential $U(\bm{\tilde\rho})=g_{FB}n_B(\bm{\tilde\rho})+V_F(\bm{\tilde\rho})$ felt by the fermions, with $V_F(\bm{\tilde\rho})=\tfrac{1}{2}\sum_i m\omega_{0i}^2\tilde\rho_i^2$ being the external trapping potential. We use the radial Thomas-Fermi radius of the condensate  $R_{\perp}=\sqrt{(2\mu-\delta)/m\omega_r^2}$ as a unit of length.}
\label{Fig1}
\end{figure}

By substituting the Thomas-Fermi solution of Eq. (10b)
\begin{equation}
g_{BB}|\chi_B(\bm \varrho)|^2=\mathcal{V}_c [2\mu-\delta-\sum_i m c_i (\omega_i^2c_i+\ddot c_i)\varrho_i^2]
\end{equation}
into Eq. (10a), making the replacement $\varrho_i=b_i/c_i\rho_i$ and imposing the relation
\begin{equation}
\ddot b_i+\omega_i^2 \left (1-\frac{2g_{FB}}{g_{BB}}\right)b_i-\frac{2g_{FB}}{g_{BB}}\frac{b_i}{c_i}\ddot c_i=\omega_{0i}^2 \left (1-\frac{2g_{FB}}{g_{BB}}\right)b_i^{-3}
\end{equation} 
on the scaling parameters $b_i(t)$, one arrives at a set of equations for the factorized fermion field operator
\begin{equation}
\label{IdealFermi}
i\hbar\frac{\partial\hat\xi_{\uparrow,\downarrow}}{\partial\tau_i}=\left[-\frac{\hbar^2}{2m}\frac{\partial^2}{\partial\rho_i^2}+U_i(\rho_i)\right]\hat\xi_{\uparrow,\downarrow}(\rho_i)
\end{equation}
with $\tau_i(t)=\int\limits^{t}b_i^{-2}(t^\prime)dt^\prime$ and
\begin{equation}
\label{EffectivePotential}
U_i(\rho_i)=\tfrac{1}{2}m\tilde\omega_{0i}^2\rho_i^2+\frac{g_{F B}}{g_{BB}}(2\mu-\delta)
\end{equation}
with $\mu$ and $\delta$ being, respectively, the chemical potential and detuning in the new units of time (multiplied by $b_i^2$). According to our convention adopted in the main text, we use the same notation for the rescaled energy units. We have defined
\begin{equation}
\tilde\omega_{0i}^2=\omega_{0i}^2 \left (1-\frac{2g_{FB}}{g_{BB}}\right).
\end{equation}
For 3D samples of ultra-cold atoms one has $g_{FB}/g_{BB}\approx 3$ \cite{Petrov2003, Petrov2004}, so that $\tilde\omega_{0i}^2<0$. In the semiclassical approximation the three equations on $\hat\xi_F(\rho_i)$ ($i=1,2,3$ and we've replaced the spin labels by "$F$") yield a simple Thomas-Fermi relation between the local Fermi energy $\epsilon_F(\bm\rho)=\hbar^2  k_F^2(\bm \rho)/2$ and the global chemical potential:
\begin{equation}
\mu=\epsilon_F(\bm\rho)+\sum_i U_i(\rho_i).
\end{equation}
The fermion density $n_F=\braket{\hat\chi_\uparrow^\dagger\hat\chi_\uparrow}=\braket{\hat\chi_\downarrow^\dagger\hat\chi_\downarrow}$ is related to the local Fermi energy by \cite{PitaevskiiBook}
\begin{equation}
\epsilon_F(\bm\rho)=\frac{\hbar^2}{2m}\left [6\pi^2 n_F (\bm\rho)\right]^{2/3}.
\end{equation}
By taking the Thomas-Fermi radius of the condensate in the radial direction $R_{\perp}=\sqrt{(2\mu-\delta)/m\omega_r^2}$ as a unit of length and introducing the aspect ratio $\sigma\equiv R_{\perp}/R_z$ with $R_z=\sqrt{(2\mu-\delta)/m\omega_z^2}$, we obtain an equation for the local Fermi energy (now in units of $\mu$) in cylindrical coordinates
\begin{equation}
\label{FermiEnergyInside}
\tilde\epsilon_F(\bm{\tilde\rho})=1-\tfrac{g_{FB}}{g_{BB}}\left (2-\frac{\delta}{\mu}\right)+\left(\tfrac{2g_{FB}}{g_{BB}}-1\right)\left (1-\frac{\delta}{2\mu}\right)(\tilde\rho^2+\sigma^2\tilde \rho_z^2),
\end{equation}
valid at 
\begin{equation*}
\frac{2\tfrac{g_{FB}}{g_{BB}}(1-\delta/2\mu)-1}{(2\tfrac{g_{FB}}{g_{BB}}-1)(1-\delta/2\mu)} \leqslant \tilde\rho^2+\sigma^2\tilde \rho_z^2\leqslant 1.
\end{equation*}
In the same units, the dimensionless equation for the effective repulsive potential produced by the condensate reads
\begin{equation}
\tilde U(\bm{\tilde\varrho})=\frac{g_{FB}n_B(\bm{\tilde\varrho})}{\mu}=\tfrac{2g_{FB}}{g_{BB}}\left (1-\frac{\delta}{2\mu}\right)(1-\tilde\varrho^2-\sigma^2\tilde \varrho_z^2),
\end{equation}
which is valid at $\tilde\varrho^2+\sigma^2\tilde \varrho_z^2\leqslant 1$. Beyond the Thomas-Fermi boundary of the condensate, at $1<\tilde\rho^2+\sigma^2\tilde \rho_z^2\leqslant (1-\delta/2\mu)^{-1}$, the fermions experience the bare external trapping potential and their local Fermi energy takes the form of an inverted parabola:
\begin{equation}
\label{FermiEnergyOutside}
\tilde\epsilon_F(\bm{\tilde\rho})=1-\left (1-\frac{\delta}{2\mu}\right)(\tilde\rho^2+\sigma^2\tilde \rho_z^2).
\end{equation}
The spatial distributions of the quantities $U(\bm\varrho)$ and $\epsilon_F(\bm{\rho})$ are presented in Fig. \ref{Fig1}.

\section{The Josephson coupling parameters}

The full second time derivative of the average squared radii for the fermionic shell ($\sigma=\uparrow,\downarrow$) has the from
\begin{equation}
\ddot W_{\sigma}^{(i)}(t)\equiv \ddot b_i(t)W_{\sigma}^{(i)}(0)=\left[\ddot W_{\sigma}^{(i)}(t)\right]^{(0)}+\frac{1}{2N_{\sigma} W_{\sigma}^{(i)}(0)}\int d\bm\rho x_i^2 \dot J_\alpha-\frac{\ddot N_\sigma}{2W_{\sigma}^{(i)}(0) N_\sigma^2}\int d\bm\rho n_\sigma x_i^2
\end{equation}
with $W_{\sigma}^{(i)}(t)\equiv \braket{\hat x_{i}^2(t)}_{\sigma}$ and
\begin{equation}
J_\alpha=\frac{2\alpha}{\hbar}\sin(\Delta\Phi)|\braket{\hat\Psi_\downarrow\hat\Psi_\uparrow}\Psi_B^\ast|.
\end{equation}
Here $\left[\ddot W_{\sigma}^{(i)}(t)\right]^{(0)}$ is due to the particle dynamics at $\alpha=0$ and we have kept only the terms that would be linear in the small variations of the scaling parameters $\delta b_i(t)$. An analogous equation for the bosonic core ($\sigma=B$) may be obtained by replacing $b_i(t)\rightarrow c_i(t)$, $\bm\rho\rightarrow \bm{\varrho}$, $N_{\uparrow,\downarrow}\rightarrow N_B$ and reversing the sign of the Josephson current $J_\alpha$. This yields the coupled equations for the scaling parameters $b_i$ and $c_i$, where the coupling elements are given by 
\begin{subequations}
\label{Couplings}
\begin{align}
\Gamma_{ij}&=\frac{\alpha m}{\hbar^2}\cos(\Delta\Phi)\left (\frac{\int d\bm\rho\rho_i^2\rho_j^2|\braket{\hat\chi_\downarrow\hat\chi_\uparrow}\chi_B^\ast|}{\int d\bm\rho\rho_i^2\braket{\hat\chi_\uparrow^\dagger\hat\chi_\uparrow}}-\frac{\int d\bm\rho\rho_j^2|\braket{\hat\chi_\downarrow\hat\chi_\uparrow}\chi_B^\ast|}{\int d\bm\rho\braket{\hat\chi_\uparrow^\dagger\hat\chi_\uparrow}}\right)\\
\Lambda_{ij}&=\frac{\alpha m}{\hbar^2}\cos(\Delta\Phi)\left (\frac{\int d\bm\varrho\varrho_i^2\varrho_j^2|\braket{\hat\chi_\downarrow\hat\chi_\uparrow}\chi_B^\ast|}{\int d\bm\varrho\varrho_i^2|\chi_B|^2}-\frac{\int d\bm\varrho\varrho_j^2|\braket{\hat\chi_\downarrow\hat\chi_\uparrow}\chi_B^\ast|}{\int d\bm\varrho |\chi_B|^2}\right),
\end{align}
\end{subequations}
where $i=j$ is admissible.

\section{Evaluation of the parameters $\Gamma(\delta)$ and $\Lambda(\delta)$ as functions of the detuning $\delta$}

By using the identities
\begin{equation*}
\begin{split}
&\int\limits_0^{2\pi}\cos^4\theta d\theta=\int\limits_0^{2\pi}\sin^4\theta d\theta=\frac{3\pi}{4}\\
&\int\limits_0^{2\pi}\sin^2\theta\cos^2\theta d\theta=\frac{\pi}{4}
\end{split}
\end{equation*}
the differences
\begin{equation*}
\begin{split}
&\Gamma(\delta)\equiv\Gamma_{xx}-\Gamma_{xy}\\
&\Lambda(\delta)\equiv\Lambda_{xx}-\Lambda_{xy}
\end{split}
\end{equation*}
may be recast in the form of Eq. (27) presented in the main text. Further evaluation of $\Gamma(\delta)$ and $\Lambda(\delta)$ as a function of $\delta$ may be performed by changing to the dimensionless variables introduced in Section I and using the expressions for the relevant spatial distributions obtained therein:
\begin{subequations}
\begin{align}
\Gamma(\tilde\delta)&=\frac{\alpha m}{\hbar^2}R_\mathrm{TF}^2\left(\frac{\hbar^2}{2m}\right)^{3/2}\frac{3\pi^2}{|\bar g_{\uparrow\downarrow}|\sqrt{g_{FB}}}\left(1-\frac{\tilde\delta}{2}\right)\frac{\int \tilde \epsilon_F(\tilde\rho)\exp\left[-\frac{\pi}{2\mathsf{k}_F\sqrt{\tilde\epsilon_F(\tilde\rho)}}\right]\sqrt{\tilde U(\tilde\rho)}\tilde\rho^5 d\tilde\rho}{\int\tilde\epsilon_F^{3/2}(\tilde\rho)\tilde\rho^3d\tilde\rho}\\
\Lambda(\tilde\delta)&=\frac{\alpha m}{2\hbar^2}R_\mathrm{TF}^2\frac{\sqrt{\mu}\sqrt{g_{FB}}}{|\bar g_{\uparrow\downarrow}|}\left(1-\frac{\tilde\delta}{2}\right)\frac{\int \tilde \epsilon_F(\tilde\rho)\exp\left[-\frac{\pi}{2\mathsf{k}_F\sqrt{\tilde\epsilon_F(\tilde\rho)}}\right]\sqrt{\tilde U(\tilde\rho)}\tilde\rho^5 d\tilde\rho}{\int\tilde U(\tilde\rho)\tilde\rho^3d\tilde\rho},
\end{align}
\end{subequations}
where $R_\mathrm{TF}^2\equiv 2\mu/m\omega_r^2$ and $\mathsf k_F\equiv \sqrt{2m\mu \bar a_{\uparrow\downarrow}^2/\hbar^2}$. We approximate our cigar-shaped cloud by an infinite tube ($\sigma=0$) to simplify the numerics. Since we here aim at order-of-magnitude estimates, the resulting minor error is of no importance. Values of the relevant parameters are listed below. We take $\Delta B=0.02$ T, $\mu_B=9.3\times 10^{-6}$ kg/s$^2\times$nm$^2$/T, $m=10\times 10^{-27}$ kg, $\bar a_{\uparrow\downarrow}=-1500a_0$, $a_0=0.05$ nm \cite{Zurn2013}. This yields $\alpha=6.62\times 10^{-6}$ kg$\times$nm$^2\times$nm$^{3/2}$/s$^2$. Following Refs. \cite{Petrov2003, Petrov2004}, we take $\bar a_{FB}=1.2 |\bar a_{\uparrow\downarrow}|$ and $\bar a_{BB}=0.6 |\bar a_{\uparrow\downarrow}|$. Note, that, by virtue of the Pauli exclusion principle, interaction of a fermion with a tightly bound molecule is always repulsive. Furthermore, we take $\mathsf k_F\equiv k_F |\bar a_{\uparrow\downarrow}| =0.3$ (in the experiment \cite{Altmeyer2007} one had $(k_F a)^{-1}=-0.85$ at $a=-5000 a_0$). This yields the chemical potential $\mu=8\times 10^{-12}$ kg$\times$nm$^2$/s$^2$, the Thomas-Fermi radius $R_\mathrm{TF}\approx 20\times 10^3$ nm and $N=1/3\mu^3/(\hbar^3\omega_r^2\omega_z)\approx 200 000$ for $\omega_r=2\pi\times270$ Hz and  $\omega_z=2\pi\times 22$ Hz.

\section{Comparison of the energy gains $E_\alpha(\delta)$ and $E_q(\delta)$}

\begin{figure}[t]
\centering
\includegraphics[width=0.8\columnwidth]{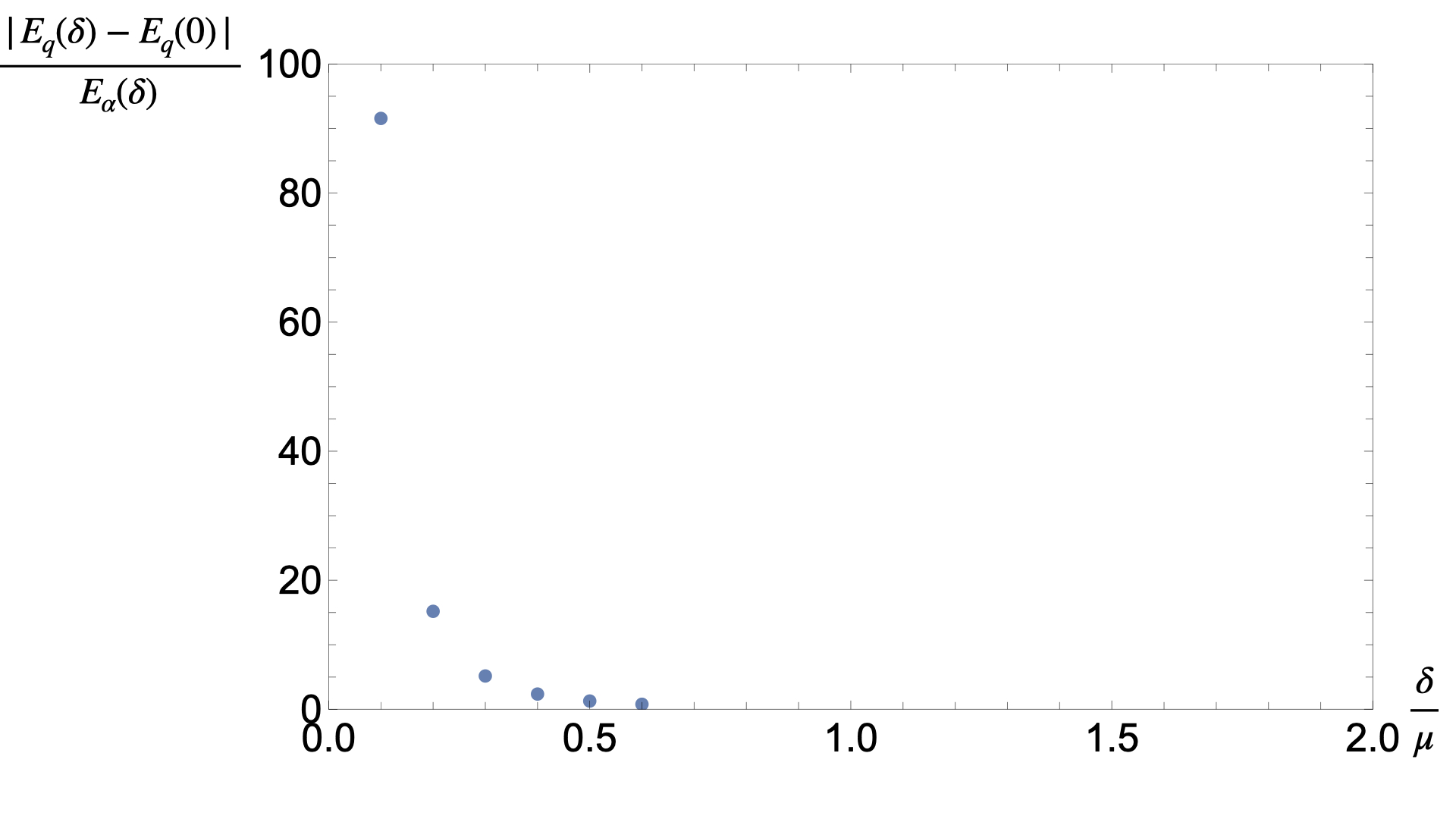}
\caption{Ratio $|E_q(\delta)-E_q(0)|/|E_\alpha(\delta)|$ as a function of $\delta$. More refined calculations are needed to certify the behaviour at $\delta>0.5\mu$.}
\label{Fig2}
\end{figure}

The energy gain due to the Feshbach coupling realized at $\Delta\Phi=0$ reads
\begin{equation}
E_\alpha(\delta)=-2\alpha\int |\braket{\hat\chi_\downarrow\hat\chi_\uparrow}\chi_B^\ast| d\bm\rho.
\end{equation}
The change of the energy associated with the quadrupole excitation can be written as
\begin{equation}
E_q(\delta)-E_q(0)=\sqrt{2}N\hbar\omega_r[\sqrt{1+\Gamma(\delta)+\Lambda(\delta)}-1].
\end{equation}
Provided $|E_q(\delta)-E_q(0)|>|E_\alpha(\delta)|$, the energy is minimized by the choice $\Delta\Phi=\pi$. In Fig. \ref{Fig2} we plot the ratio $|E_q(\delta)-E_q(0)|/|E_\alpha(\delta)|$ as a function of $\delta$. We use the parameters listed in Section III. Besides, we now have to retain finite value of the aspect ration $\sigma=\omega_z/\omega_r$. One can see that the ratio greatly exceeds unity at small $\delta$. However, it drops quickly as $\delta$ in increased. This may potentially result in switching of $\Delta\Phi$ back to $0$. More refined studies are needed to address this issue.

\bibliography{Bibliography}